\def\year{2020}\relax
\documentclass[letterpaper]{article} 
\usepackage{aaai20}  
\usepackage{times}  
\usepackage{helvet} 
\usepackage{courier}  
\usepackage[hyphens]{url}  
\usepackage{graphicx} 
\usepackage{graphicx}  
\usepackage{amssymb}
\usepackage{amsmath}
\usepackage{setspace}
\DeclareMathOperator*{\argmax}{arg\,max}

\title{TableQA: a Large-Scale Chinese Text-to-SQL Dataset for \\
Table-Aware SQL Generation}

\author{
  Ningyuan Sun,
  Xuefeng Yang,
  Yunfeng Liu
  \\\Large{Zhuiyi Technology}\\
  {\tt \large{\{waynesun, ryan, glenliu\}@wezhuiyi.com}}\\
}
\begin{document}
\maketitle
\begin{abstract}
Parsing natural language to corresponding SQL (NL2SQL) with data driven approaches like deep neural networks attracts much attention in recent years. Existing NL2SQL datasets assume that condition values should appear exactly in natural language questions and the queries are answerable given the table. However, these assumptions may fail in practical scenarios, because user may use different expressions for the same content in the table, and query information outside the table without the full picture of contents in table. Therefore we present TableQA, a large-scale cross-domain Natural Language to SQL dataset in Chinese language consisting 64,891 questions and 20,311 unique SQL queries on over 6,000 tables. Different from exisiting NL2SQL datasets, TableQA requires to generalize well not only to SQL skeletons of different questions and table schemas, but also to the various expressions for condition values. Experiment results show that the state-of-the-art model with 95.1\% condition value accuracy on WikiSQL only gets 46.8\% condition value accuracy and 43.0\% logic form accuracy on TableQA, indicating the proposed dataset is challenging and necessary to handle. Two table-aware approaches are proposed to alleviate the problem, the end-to-end approaches obtains 51.3\% and 47.4\% accuracy on the condition value and logic form tasks, with improvement of 4.7\% and 3.4\% respectively.
\end{abstract}

\section{Introduction}

Semantic parsing aims to map the semantic meaning of natural language to interpretable logic representations, which has been applied in many fields such as translation\cite{shvets2019improving}, question answering\cite{cheng2016long}, and robot navigation\cite{kwiatkowski2013scaling}. Parsing Natural Language to Structured Query Language (NL2SQL) is one of the most typical tasks, which has been under study for a long period of time. The early remarkable work may be traced back to the LUNAR system\cite{Woods:1973:PNL:1499586.1499695} developed in the 1970s. However, limited by the scale of datasets and computing power, early solutions mainly relied on techniques like pattern matching, syntax tree and semantic grammar. 
\begin{figure}[!t]
    \hspace{-1mm}
    \centering
    \includegraphics[width=1\columnwidth]{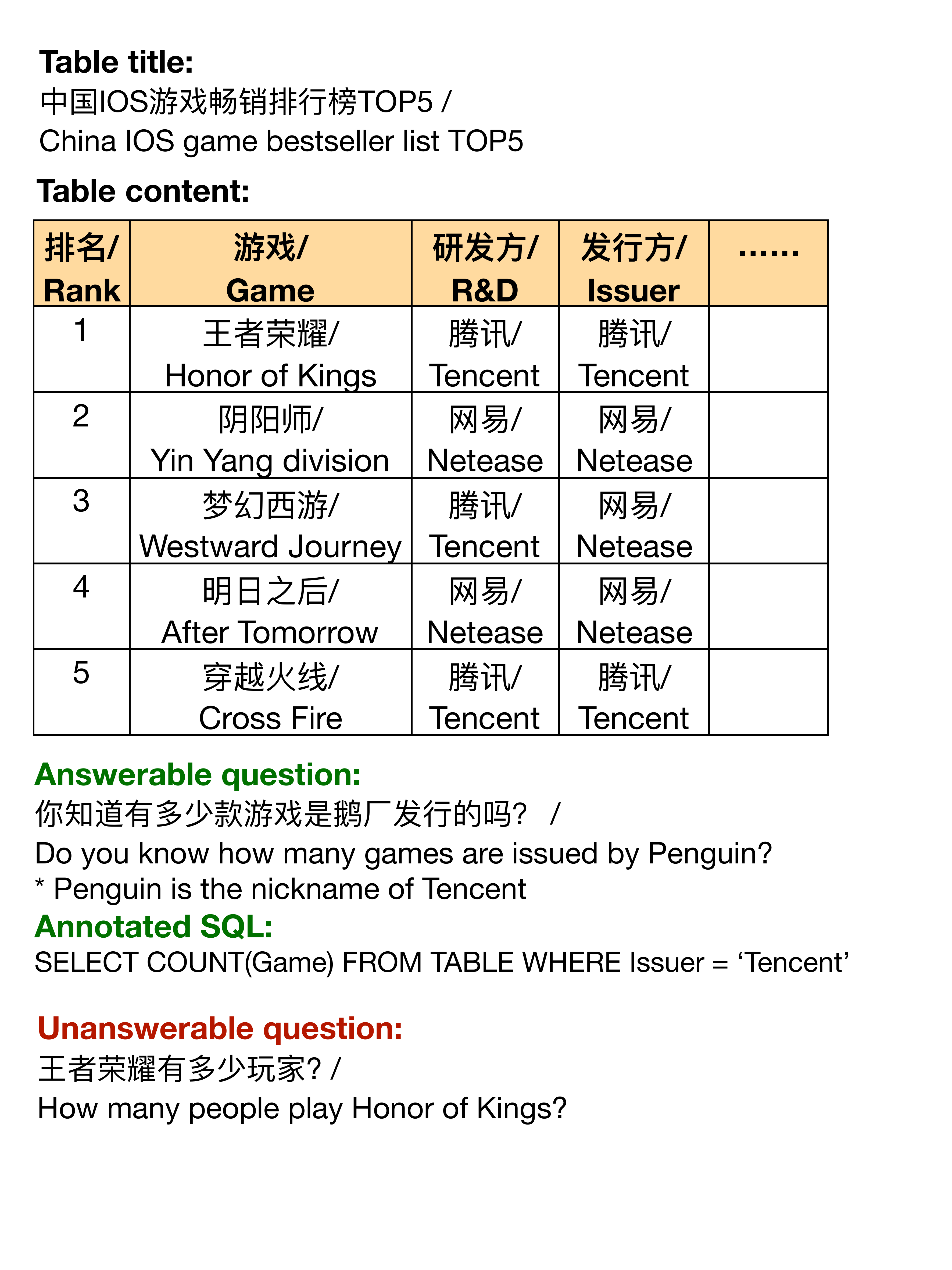}
    \caption{
    The samples in TableQA contain answerable question and the corresponding SQL program based on given table. Topically related unanswerable questions are also included. 
    }
\label{figure_sample_data}
\end{figure}

With the development of deep neural network and data-driven methods, the release of WikiSQL\cite{zhongSeq2SQL2017} brings NL2SQL to a new era. 80,654 natural language question and SQL pairs make deep learning applicable. Spider\cite{Yu&al.18c} is another typical large-scale NL2SQL dataset. Compared with WikiSQL, Spider defines the NL2SQL problem in a more challenging setting with complex SQL skeleton and database schema. 

Although WikiSQL and Spider are well designed and make remarkable contributions to the development of end-to-end solutions for NL2SQL, two simplified but necessary problems are ignored in both of them, including entity linking and answerability. WikiSQL and Spider assume that users' questions are expressed with the same mentions as in the tables, which is not realistic due to the fact that there are different kinds of expressions for the same entity or concept. Through our statistics, over 97\% and 95\% of the condition values in SQLs appear in the questions exactly the same in WikiSQL and Spider respectively. This supports the fact that WikiSQL and Spider ignore entity linking. Meanwhile, all questions are sure to find the answer according to the given table. However, topically related but unanswerable questions may occur when the table does not contain the information queried by users, it is reasonable and necessary to tell users directly whether the table contains the information being queried. 

To address entity linking, answerability and NL2SQL jointly, we present TableQA, a large scale cross domain NL2SQL dataset in Chinese. It consists of 64,891 questions and 20,311 unique SQL queries on over 6,000 tables from various domains. Among the annotated questions, over 5,000 questions are unanswerable according to the given table. Figure \ref{figure_sample_data} shows one sample data from TableQA as example, including a table, an answerable question and an unanswerable question. 

To the best of our knowledge, TableQA is the first NL2SQL dataset to bring entity linking and answerabiltiy together. In semantic parsing area, there has been datasets involving these elements, however, some differences exist between TableQA and others. We collect several popular question answering datasets\cite{QALD,WebQuestionsSP,LCQuad}. Comparison on entity linking, answerabilty, output language and scale aspects is shown in Table 1.

In addition, we adopt two table-aware baseline approaches for the entity linking and answerability problems. The first approach adds an offline step to recognize the value for each condition from the corresponding column, and the second solution employs an table-aware mechanism to predict condition value. Both of the methods make significant improvements on the basis of existing solutions. 

\section{Existing Dataset and Related Work}\label{related_work}
Decades’ studies of semantic parsing have accumulated lots of valuable datasets. ATIS\cite{data-atis-original} and GeoQuery\cite{data-geography-original} contain complex SQL in flight and geography domains respectively. Scholar\cite{data-atis-geography-scholar} and Academic\cite{Academic} are specific in academic area, with different schemas. Other datasets include Jobs\cite{tang2001using}, Advising\cite{data-sql-advising}, Yelp and Imdb\cite{data-sql-imdb-yelp}, etc. Besides research article, practical application systems are developed like SimpleQL\cite{kueri}.

\begin{table} [!tb]
  \centering
  \label{compare_dataset}
    \begin{tabular}{c|cccccc} 
      \hline 
      Dataset & EL & Answer & Language & Scale\\ 
	  \hline 
	  SimpleQ & \checkmark &  $\times$ & $\times$ & 10,000\\
	  WebQ & \checkmark &  $\times$ & $\times$ & 5,810\\
	  WebQSP & \checkmark &  $\times$ & SPARQL & 4,737\\
	  LC-Quad & \checkmark & \checkmark & SPARQL & 5,000\\
	  QALD & \checkmark & \checkmark & SPARQL & 408\\
	  WikiTableQ & \checkmark & \checkmark & $\times$ & 22,033 \\
	  WikiSQL & $\times$ & $\times$ & SQL & 80,654 \\
	  Spider & $\times$ & $\times$ & SQL & 10,181 \\ 
	  \hline
	  TableQA & \checkmark & \checkmark & SQL & 64,891 \\
      \hline 
    \end{tabular}
    \caption{Comparisons on features between existing semantic parsing question answering datasets, including SimpleQuestions, WebQuestions, WebQuestionsSP, LC-Quad, QALD, WikiTableQuestions, etc. EL, Answer, Langauge, Scale represent whether involves entity linking, answerability, output language and datatset scale respectively}
\end{table}

Recently, deep learning approaches are widely applied in semantic parsing and obtaining promising results. Three large scale datasets suitable for data-driven approaches are created. WikiSQL is one of the datasets, which contains 80,654 questions based on 24,241 tables extracted from Wikipedia, and supported SQL skeletons are limited but practical. Spider consists of 10,181 questions and 5,693 unique SQLs, and it is the first large scale NL2SQL dataset supporting multi tables joint queries. Different from WikiSQL and Spider with correct SQL as training label, WikiTableQuestions\cite{pasupat2015compositional} is a question answering dataset based on semi-structured table without specific logic form, taking the answer of question as corresponding label. Entity linking was also introduced in WikiTableQuestions, over 20\% questions are not answerable by existing execution models, and neural based method is proposed to handle these challenges\cite{krishnamurthy-etal-2017-neural}. However, standard SQLs are not provided in the dataset, which is different from TableQA with correct SQLs. Besides, the unanswerable questions are caused by unhandled question types or annotation mistakes, instead of intentionally design in TableQA. Similar situations occur in other semantic parsing datasets\cite{WebQuestions,SimpleQuestions}. Therefore, TableQA raises new form but practical challenges in NL2SQL area.
\begin{figure*}[!t]
    \centering
    \includegraphics[width=1\textwidth]{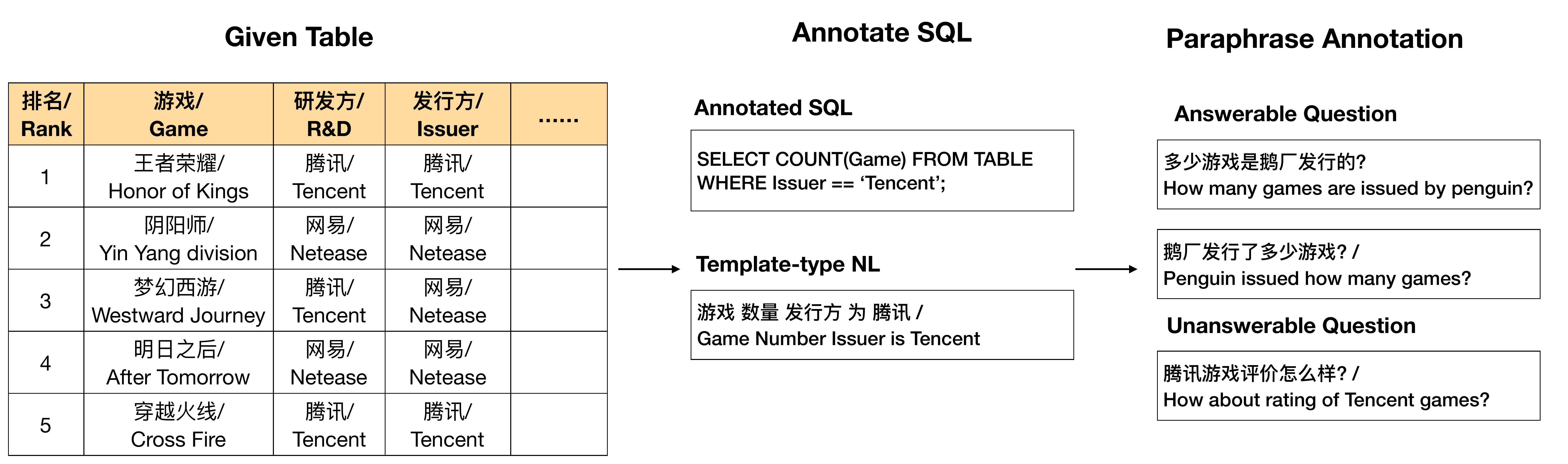}
    \caption{
    Brief description of data annotation process.
    }
\label{Annotation_process}
\end{figure*}

These NL2SQL datasets inspire deep learning methods, which could be grouped into two categories according to problem definition and model structures. One category employs seq2seq model structure and defines the problem as a sequential generation problem like translation, like MQAN\cite{mccann2018natural}. The another category decouples the the SQL generation into several subtasks based on SQL skeleton, and usually perform better than sequential generation methods under simple SQL skeleton, including X-SQL\cite{xsql}, SQLova\cite{hwang2019achieving}, SQLNet\cite{xu2017sqlnet}. Among these solutions, X-SQL has already achieved 91.8\% execution accuracy on WikiSQL. However, these datasets pay much attention to the scale of dataset or emphasize complexity in SQL skeletons, ignoring to combine natural language semantics and table content while generating SQL. \cite{sun2018semantic} proposed a table-aware method to generate SQL using the dataset of WikiSQL, but only about 3\% of condition values do not appear in natural language utterance in WikiSQL, this proportion is not sufficient to emphasize the problem of entity linking.

TableQA is designed with practical considerations. To be specific, annotators are encouraged to propose questions with different entities or other terms from the ones stored in tables. Meanwhile, different from other datasets, some questions may be not answerable according to the given table, due to lack of enough information. These two factors make TableQA more close to the real scenario than other existing NL2SQL datasets.  

\section{TableQA}\label{dataset}
To discover the entity linking and answerability problems under the NL2SQL scenario, we propose a NL2SQL dataset named TableQA. The building process is divided into three phases. Firstly, the tables used for annotation are crawled from public web resources and financial report. Then, the annotation of natural language question and SQL pairs are performed according to the collected tables. Finally, the annotated samples are verified manually to guarantee the quality of data. 

\subsection{Principles of Annotation}\label{principles}
Before introducing the details of annotation, we first elaborate the principles of designing this dataset. To emphasize entity linking problem, annotators are required to follow three principles while annotating samples, including expression diversity, schema omission and unanswerable question. Besides the above requirement, the sample size should be large enough to support the data driven approaches like deep neural network. The  included SQL skeletons is another point to be considered, and finally, tables in cross domains are employed to test generalization ability of learned systems.

\subsubsection{Expression Diversity}
For part of the data, their ground truth of condition values should not appear in natural language questions exactly. In practical scenarios, users may not use the exactly same word as the data saved in table when they ask question about an entity, term or concept. For example, a city name stored in table may be \textit{Los Angeles}, but users may express question using \textit{LA}, other examples such as \textit{return on equity} as \textit{ROE}, \textit{Tencent} as \textit{penguin}, etc. Therefore, annotators are encouraged to paraphrase the questions for the same intention, especially for the condition value. Moreover, expression diversity covers not only entities, but also predicates, adjectives and so on. More details may be illustrated in \ref{dataset_statistics}

\subsubsection{Schema Omission}
Annotators are encouraged to annotate the natural language question in a natural format. For example, compared with ``What is the stock price of the company named Salesforce?", users tend to express ``What is the stock price of Salesforce?". The latter expression hide the strong feature of \textit{company name}, which requires the system to infer that the column related to condition should be the \textit{company name}.

\subsubsection{Unanswerable Question}
It is quite likely that the answers to users' question cannot be found in tables. This phenomenon is called ``Empty prompt” or ``Habitability problem" in relevant literature and the conclusion is that users should be informed the reason why the return is none to avoid confusion. Therefore, annotators of TableQA are encouraged to design some questions that could not be answered given the corresponding table, though they are topically related. This setting makes the execution guide decoding(EGD) \cite{EGD} not applicable, which is widely employed to enhance performance for WikiSQL dataset. To be more specific, the model with EGD may try the other high-probable prediction if the execution of predicated SQL return exceptions or empty set, this procedure is iterated until the SQL execution result is not none.

\subsection{Table Collection}
Two kinds of resources are employed to collect the tables. The first source is various kinds of financial reports. Financial reports contain a large number of valuable tables, like balance sheet, financial analysis, industrial analysis, etc. Over 1,500 financial reports are collected from public website, and the tables are filtered out from these reports. 

Another source is a set of spreadsheet files collected through Google. Firstly, 10,000 frequently used words are summarized from the Baidu Baike corpus, including words like expense, ticket, shop, food. Then these words are queried in Google and filtered spreadsheet files are downloaded. Useful tables are extracted from these files through a parser, which could identify potential table in a worksheet. 

\subsection{Data Annotation}
A simple annotation platform is developed for annotating natural language questions and corresponding SQL. Annotators are provided with a full view of specific sampled from the tables collected, which contains the table name, column names, table content and column data types, e.g. number, text. After comprehending the given table, annotators propose reasonable questions according to the table. Instead of typing SQL manually, annotators create SQL programs by selecting through an interactive user interface, and then the SQL program is generated automatically. For the created SQL program, there is also a compatible natural language sentence which is generated based on templates by annotation tool. Finally, annotators paraphrase the sentence into 2 or 3 colloquial answerable questions, and annotate 1 or 2 unanswerable sentences. A brief description of data annotation process is shown in Figure \ref{Annotation_process}.

\subsection{Data Review}
Data annotation process is performed by ten persons for three months, and cross review is conducted simultaneously, which means the fault data may be detected and corrected by other annotators. The most important point to check is whether the paraphrase procedure reserves the meaning in the initial SQL template. Meanwhile, distributions of questions on various aspects are monitored, including ratio of unanswerable questions in all samples, proportion of samples that require entity linking, distribution of aggregation function, distribution of conditions, etc. 

\begin{figure*}[!t]
    \centering
    \includegraphics[width=1\textwidth]{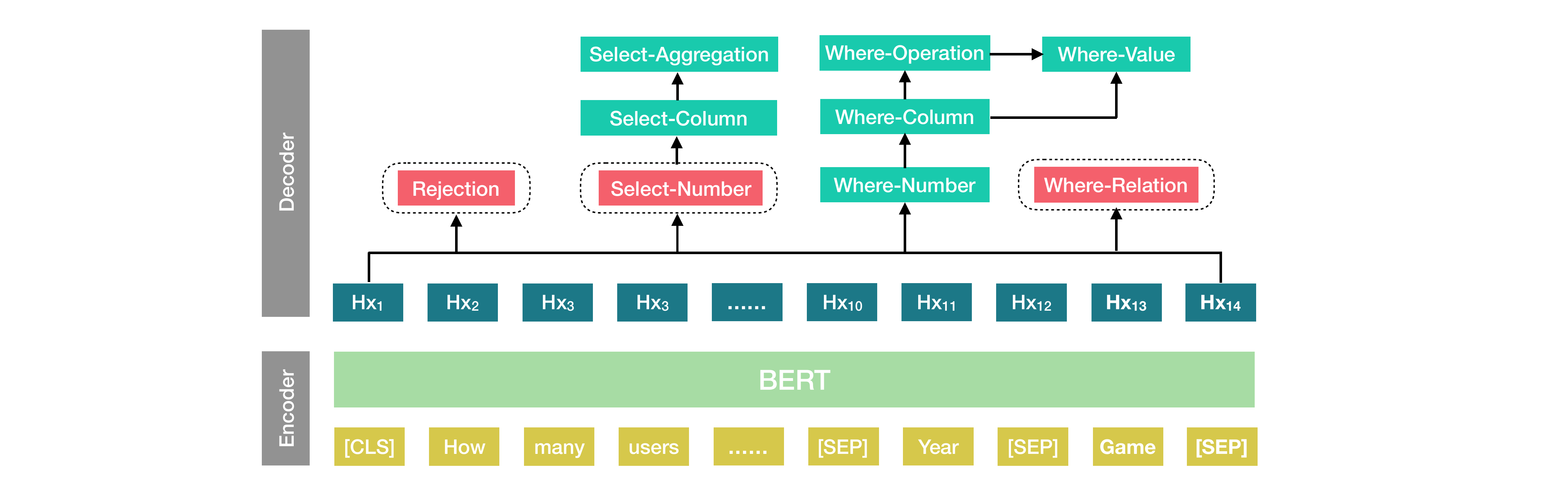}
    \caption{
    Our neural network model architecture based on SQLova. The circle parts are newly added subtasks.
    }
\label{model_structure}
\end{figure*}

\subsection{Dataset Statistics}\label{dataset_statistics}
Comparison statistics between WikiSQL, Spider and other text-to-SQL datasets on scale is presented in Table 2. As for the data scale, TableQA contains 64,891 questions, 20,311 unique SQL and over 6,000 tables, which is close to WikiSQL. Generally, TableQA satisfies nearly all major advantages of previous datasets, e.g. having large size of examples, containing complex SQL skeletons and tables from various domains. As expected, TableQA presents two unique characteristics, including entity linking and answerability, which present new challenges to the semantic parsing research community.

To give a more comprehensive analysis on TableQA, statistics from different angles is provided.  

\subsubsection{Diversity}
More than 30\% of samples need to be solved through entity linking, while this ratio becomes 3\% and 5\% for WikiSQL and Spider. This part of data could be divided into several categories, including abbreviation, alias, inconsistent number format, adaptation and others, accounting for 20.1\%, 14.1\%, 25.7\%, 32.5\% and 7.6\% through sampling statistics. 

\subsubsection{Complexity}
In TableQA, each question corresponds to one single table, multi-table query or nested query are not involved. Supported aggregation function includes \textit{MIN}, \textit{MAX}, \textit{AVG}, \textit{SUM}, \textit{MIN} and \textit{COUNT}, and condition operator includes \textit{$>$}, \textit{$<$}, $==$ and $!=$. Each table contains 10.7 natural language questions, 7.2 columns and 41.5 rows. While each SQL contains 1.6 conditions and 1.1 selected columns on average. The average lengths of natural language questions and SQL are 26.4 and 11.0 respectively.

\section{Table-aware Approaches}\label{methods}
Unlike previous dataset assuming the condition value expressions are consistent in questions and tables, TableQA presents the value normalization problem. To tackle this new challenge, two table-aware approaches are proposed based on existing end-to-end solution. The first one is an offline method which works as a post-processing step compatible with most end-to-end solutions. Another method follows the main structure of SQLova and models the selection of condition values by attention mechanism.  
\begin{table} [!tb]
  \centering
  \label{statistics}
    \begin{tabular}{c|cccc} 
      \hline 
      Dataset & Question & SQL & Table \\
	  \hline 
	  ATIS & 5,280& 947 & 32 \\
	  GeoQuery & 877 & 247 & 6\\
	  Scholar & 817 & 193 & 7 \\
	  Academic & 196 & 185 & 15 \\
	  Yelp & 128 & 110 & 7 \\
	  Advising & 3,898 & 208 & 18 \\
	  WikiSQL & 80,654 & 77,840 & 26,531 \\
	  Spider & 10,181 & 5,693 & 4,504 \\ 
	  \hline
	  TableQA & 64,891 & 20,311 & 6,029 \\
      \hline 
    \end{tabular}
    \caption{Comparisons on of existing NL2SQL datasets. }
\end{table}

Figure \ref{model_structure} shows the model structure of SQLova, which is built based on BERT. Questions and table column names are fed into the encoder module and outputs of BERT for these inputs are utilized as the representation of questions and column names. There are 6 sub-model designed to deal with the corresponding subtasks in WikiSQL. To adapt SQLova for TableQA, another three subtasks are added, including Select-Number, Where-Relation and Rejection.     

\begin{figure}[!t]
    \hspace{-1mm}
    \centering
    \includegraphics[width=1\columnwidth]{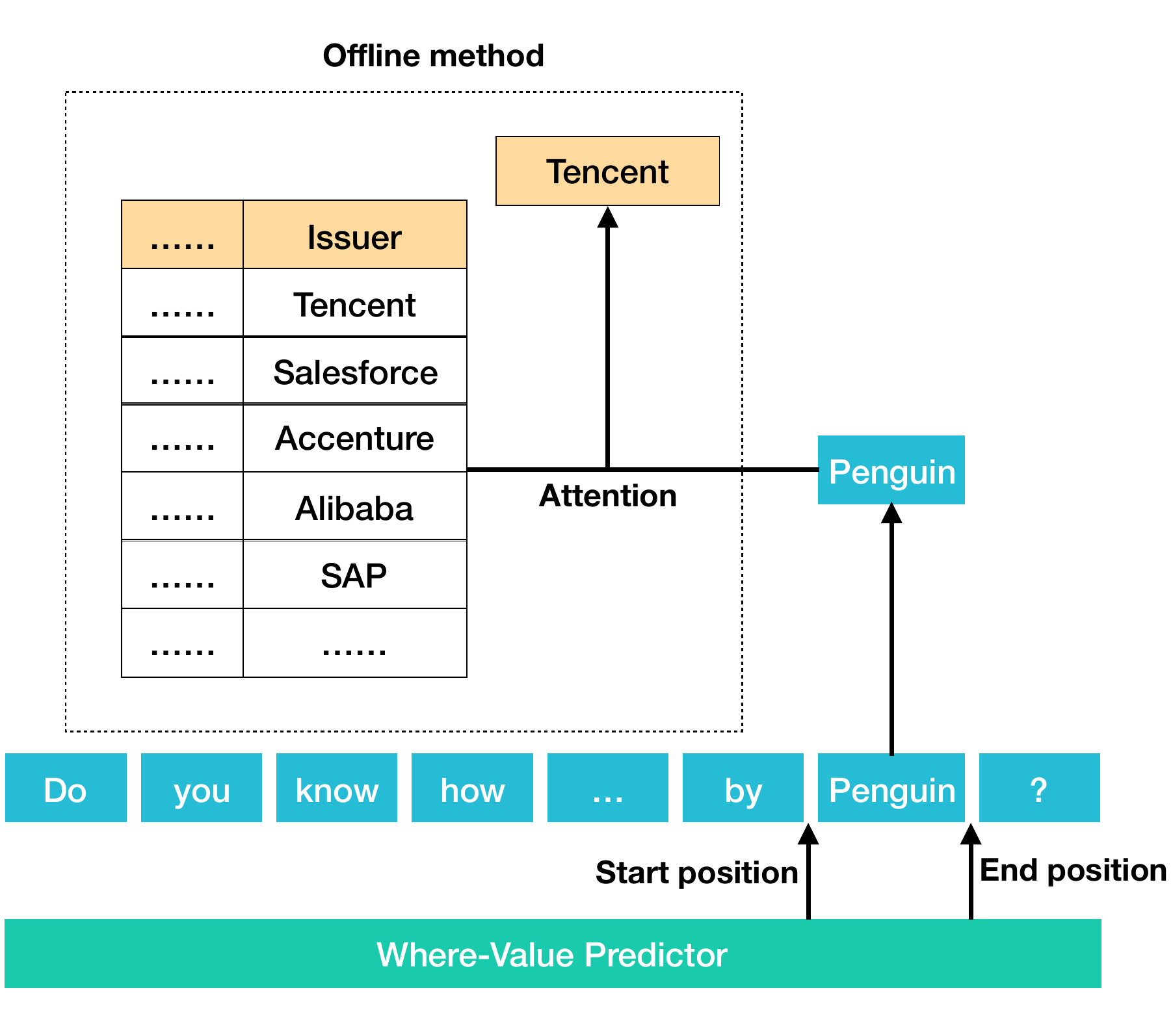}
    \caption{
    Structure of the offline method.
    }
\label{offline_method_complex}
\end{figure}

\subsection{Offline Method}
The procedure of this offline method is shown in Figure \ref{offline_method_complex}. The offline method is defined compared with the end-to-end method, which means the operation is based on the result of another end-to-end model. Firstly, a well-trained SQLova is utilized to predict the where-value of sample data, in which the condition value is a sub-string taken from original question. The start and end position indices are predicted by a pointer network \cite{pointer}, depending on the representation of the predicted where-column, predicted where-operation and embedding of the natural language question. The calculation is described as follow. 

\begin{center}
$H_n=[W_{col}H_{col_n}, W_{op}H_{op_n}, W_{att_q}H_{att_q}, H_q]$ 
\end{center}

where $n$ denotes the index of the conditioned column. $W_{col}$, $W_{op}$ and $W_{att_q} \in \mathbb{R}^{{d}\times{d}}$ are trainable variables. $H_{col_n}$, $H_{op_n}$, $H_q$ and $H_{att_q}$ denote the representation of the conditioned column, predicted condition operation, question and attended question respectively 

\begin{center}
$s_{start,n}= tanh(H_nU_{start})W_{start}$ \\
$p_{start,n}= Softmax(s_{start,n})$ 
\end{center}

where $U_{start} \in \mathbb{R}^{{4d}\times{d}}$ and $W_{start} \in \mathbb{R}^{{d}\times{d}}$ are trainable parameters.

The computation process of the end index is similar
\begin{center}
$s_{end,n}= tanh(H_nU_{end})W_{end}$ \\ 
$p_{end,n}= Softmax(s_{end,n})$
\end{center}

The representation of the predicted sub-string and each cell in the table is the average of character embeddings. This is the most commonly used baseline for vector representation of varied length text. 

\begin{center}
\begin{align}
h_{cell}= \frac{1}{N} \sum_{n=0}^Nemb_n
\end{align}
\end{center}

where $N$ denotes the length of cell or substring. Dot product is used to compute the similarity between the representations of the sub-string and the cells, and value of the cell with smallest distance is chosen. 

\begin{center}
\begin{align}
cell =  \argmax(h_{substr} \cdot h_{cell_i})
\end{align}
\end{center}

Now two candidate answers are obtained, including the sub-string extracted from question and a certain cell value in the table. Finally, the choice between two candidate answers is determined by the type of the chosen column, if data type of the corresponding column is text, then takes cell as condition value, otherwise adopts extracted sub-string.

\subsection{End-to-end Method}
Similar to the offline method, the end-to-end method is designed based on SQLova as well. For the condition value subtask, the start and end position of condition value are predicted through a pointer network. The difference is, the representation of each cell is the output of an bi-LSTM encoder with BERT\cite{BERT} lexicon embeddings as inputs, and the bi-LSTM is shared for all cells and trained respected to the loss of model. 

\begin{center}
$H^t_{cell_{n,i}}=LSTM(emb^t_{cell_{n,i}},H^{t-1}_{cell_{n,i}})$
\end{center}

\begin{table*} [ht]\label{table_3}
  \centering
  \label{experiment1}
  \resizebox{\textwidth}{13mm}{
    \begin{tabular}{c|c|c|c|cccc|c|c} 
      \hline
      Dataset & Model & S-Col & S-Agg & W-Num & W-Col & W-Op & W-Value & Logic Form & Execution \\
	  \hline
      WikiSQL & SQLNet & 91.5 & 90.1 & \multicolumn{4}{c|}{74.1} & 63.2 & 69.8 \\
	  WikiSQL & SQLova & 96.8 & 90.3 & 98.4 & 93.8 & 97.0 & 95.1 & 79.9 & 85.9 \\
	  TableQA & SQLNet & 91.5 & 93.7 & 90.1 & 71.2 & 86.5 & 51.2 & 30.1 & 34.5 \\
	  TableQA & SQLova & 96.1 & 98.1 & 95.7 & 79.4 & 93.3 & 54.3 & 43.0 & 49.7 \\
      \hline
    \end{tabular}}
    \caption{Performance of SQLNet and SQLova with base BERT on WikiSQL and TableQA respectively. The 74.1 accuracy of SQLNet represents its whole subtask of where part.}
\end{table*}

\begin{table*}[ht]
  \centering
  \label{experiment2}
  \resizebox{\textwidth}{13mm}{
    \begin{tabular}{cccccccccc} 
      \hline
      Model & S-Num & S-Col & S-Agg & W-Num & W-Col & W-Op & W-Value & W-R & Logic Form\\
	  \hline 
	  SQLNet & 98.6 & 91.5 & 93.7 & 90.1 & 71.2 & 86.5 & 43.2 & 91.2 & 30.1  \\
	  SQLova & 99.3 & 96.1 & 98.1 & 95.7 & 79.4 & 93.3 & 44.1 & 95.9 & 43.0 \\
	  SQLova+Offline & 99.3 & 96.1 & 98.1 & 95.7 & 79.4 & 93.3 & 47.4 & 95.9 & 46.6 \\
	  SQLova+End2End & 99.2 & 96.2 & 98.2 & 95.1 & 79.2 & 93.1 & 49.5 & 95.2 & 48.3 \\
      \hline
    \end{tabular}}
    \caption{Experiment results of SQLNet and SQLova on TableQA. The subtasks S-Num, S-Col, S-Agg, W-NUM, W-Col, W-Op, W-Value, W-R represent Select-Number, Select-Column, Select-Aggregation, Where-Number, Where-Column, Where-Operation, Where-Value and Where-Relationship respectively. Existing models cannot perform well on Where-Column and Where-Value task.}
\end{table*}

where $t$, $n$ and $i$ denotes time step, index of column and index of row respectively. $H^t_{cell_{n,i}}$ represents the $t$th LSTM encoder output state of the $i$th row in the $n$th column, and $emb^t_{cell_{n,i}}$ represents the embedding of the $t$th token. The last output hidden state is taken to be $H_{n,i}$, as the representation of the cell $H_{cell_{n,i}}$.

\begin{center}
$p_{cell_{n,i}}=softmax(H_nW_{row}H_{cell_{n,i}})$
\end{center}

where $W_{row}$ $\in \mathbb{R}^{{d}\times{1}}$ is a trainable variable. We use cross entropy loss for this table-aware entity linking task, and other technical details could be found in report of SQLova. 

Through this table-aware mechanism, some cases that need entity linking could be solved. To have a better understand the improvements brought by the mechanism, prediction results of a sample data in valid set is shown Figure \ref{badcase}. With the given table and question, SQLova predicts \textit{reliable} as condition value through pointer network, which is a reasonable but incorrect sub-string for SQL. After adapting the table-aware mechanism, model could predict the correct \textit{qualified} as condition value. 

\section{Experiments}\label{experiments}
Two groups of experiments are conducted to evaluate effectiveness. At first, we test whether existing state-of-the-art and baseline models could work well on entity linking and answerabiltiy in TableQA. The second experiment aims to evaluate the table-aware approaches proposed in Section \ref{methods}. 

\subsection{Experiment Setting}
\subsubsection{Existing models}
SQLNet and SQLova are utilized in our experiments. SQLNet is a baseline model used in recent NL2SQL research, while SQLova gets a state-of-the-art result on WikiSQL. Both SQLNet and SQLova decouple the problem of generating SQL into several subtasks, including Select-Column, Select-Aggregation, Where-Number, Where-Column, Where-Operation and Where-Value. Specifically, in the Where-Value subtask, a pair of start and end position in utterance is predicted through a pointer network, and the extracted sub-string is taken as condition value.

For fair comparison, these models are modified to adapt to TableQA. In detail, four new subtasks are added, including number of selected columns, aggregations for multiple selected columns, relationship between conditions and whether the question is answerable according to the given table. 

\subsubsection{Dataset Setting}
TableQA contains 64,891 questions based on over 6,000 tables, among them, over 5,500 questions are unanswerable. Ttraining, valid and test sets are distributed with 51.7K, 6.4K and 6.7K questions respectively, the part of unanswerable questions also follow the same distribution. For testing generalization ability, the tables in valid set and test set do not appear in the train set. 

\subsubsection{Evaluation Metric}
Following WikiSQL and Spider, logic form accuracy and execution accuracy are utilized as the evaluation metrics. Logic form accuracy represents the percentage of samples in which each component of the predicted SQL is exactly correct, and the execution accuracy represents the percentage of samples that get the same return as ground truth after executing SQL.
\begin{table} [!tb]
  \centering
  \label{rejection}
    \begin{tabular}{cc} 
      \hline 
      Model & F1 Score of Ans.\\
	  \hline 
	  SQLNet & 0.54 \\
	  SQLova & 0.63 \\
      \hline 
    \end{tabular}
    \caption{F1 score of existing NL2SQL models on answerability in TableQA.}
\end{table}

\subsection{Experiment Results}
In Table 3, performances of SQLNet and SQLova on WikiSQL and TableQA are presented. SQLNet achieves 63.2\% on logic form and 69.8\% on execution accuracy respectively on WikiSQL, while SQLova with base BERT\cite{BERT} may obtain 79.9\% on logic form and 85.9\% execution accuracy respectively on WikiSQL. These two results are similar to the performance reported in origin paper. However, SQLNet and SQLova only get 30.1\% and 43.0\% logic form accuracy on TableQA respectively. The accuracy drops significantly compared with their performance on WikiSQL, especially for Where-Column and Where-Value tasks. These experiment results demonstrate that TableQA is a challenging dataset for existing methods, and the major bottleneck is related to the problem of condition value. Table 5 shows the performance of exisitng models on answerability, which is measured by F1 score. Both models only achieve around 0.6, indicating existing methods are not capable of recognizing unanswerable questions, and there is still a large room for improvement. 

\begin{figure}[!t]
    \hspace{-1mm}
    \centering
    \includegraphics[width=1\columnwidth]{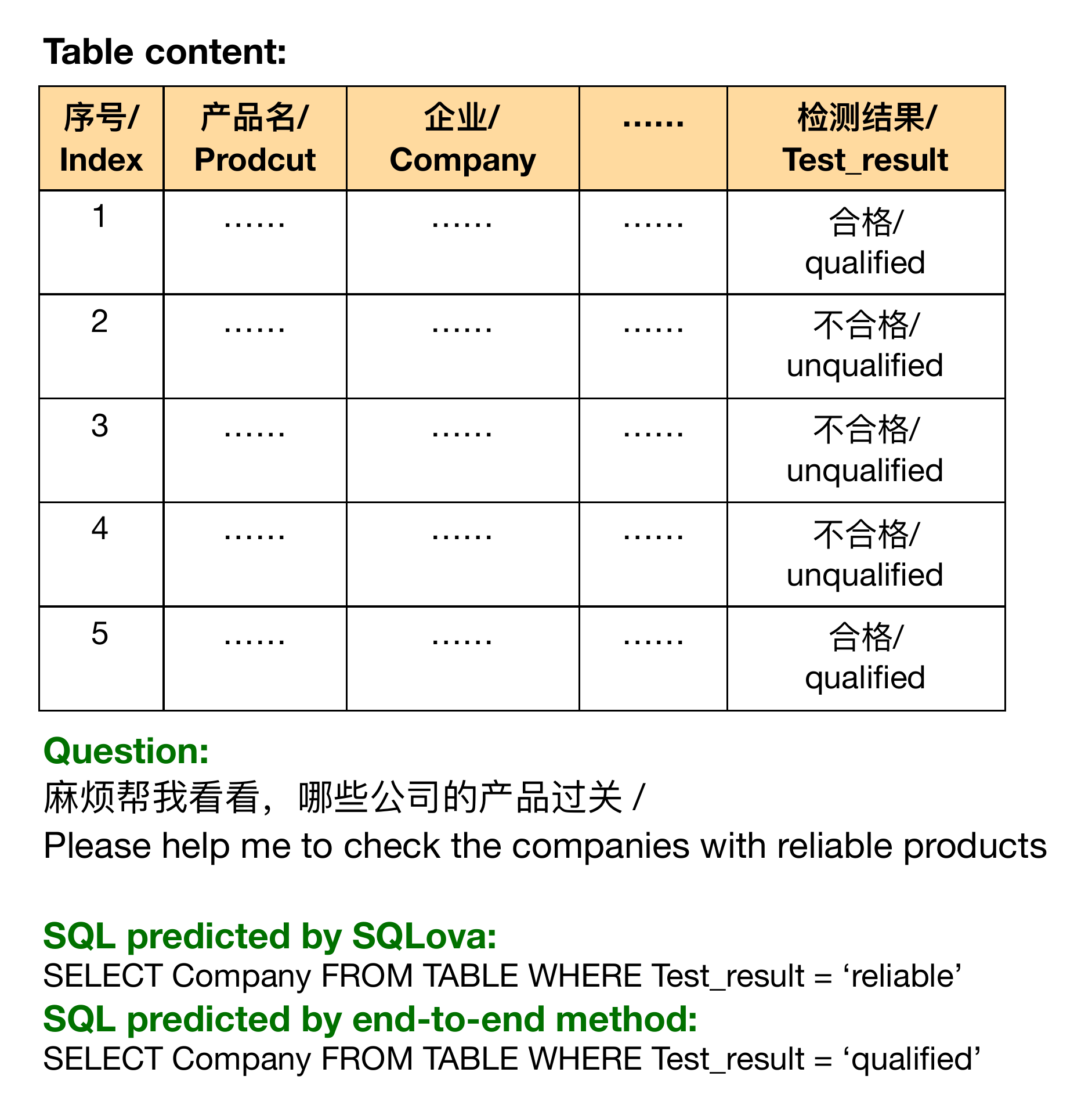}
    \caption{
    According to the given table and question, SQLova extracts a sub-string extracted from the question as condition value, which does not match with table content. After applying end-to-end table-aware mechanism, table content could be considered as condition value.  
    }
\label{badcase}
\end{figure}

The experiment result in Table 4 indicates that our proposed table-aware solutions are effective in predicting condition value. Specifically, the end-to-end model achieves 5.4\% improvement in the Where-Value subtask, compared to 44.1\% of SQLova. A detail analysis is shown in Figure \ref{badcase}.

\subsection{Error Analysis}
Experiments results indicate that TableQA is a challenging dataset. Badcases are analysed, and some interesting insights are found and summarized below.  

\begin{figure}[!t]
    \hspace{-1mm}
    \centering
    \includegraphics[width=1\columnwidth]{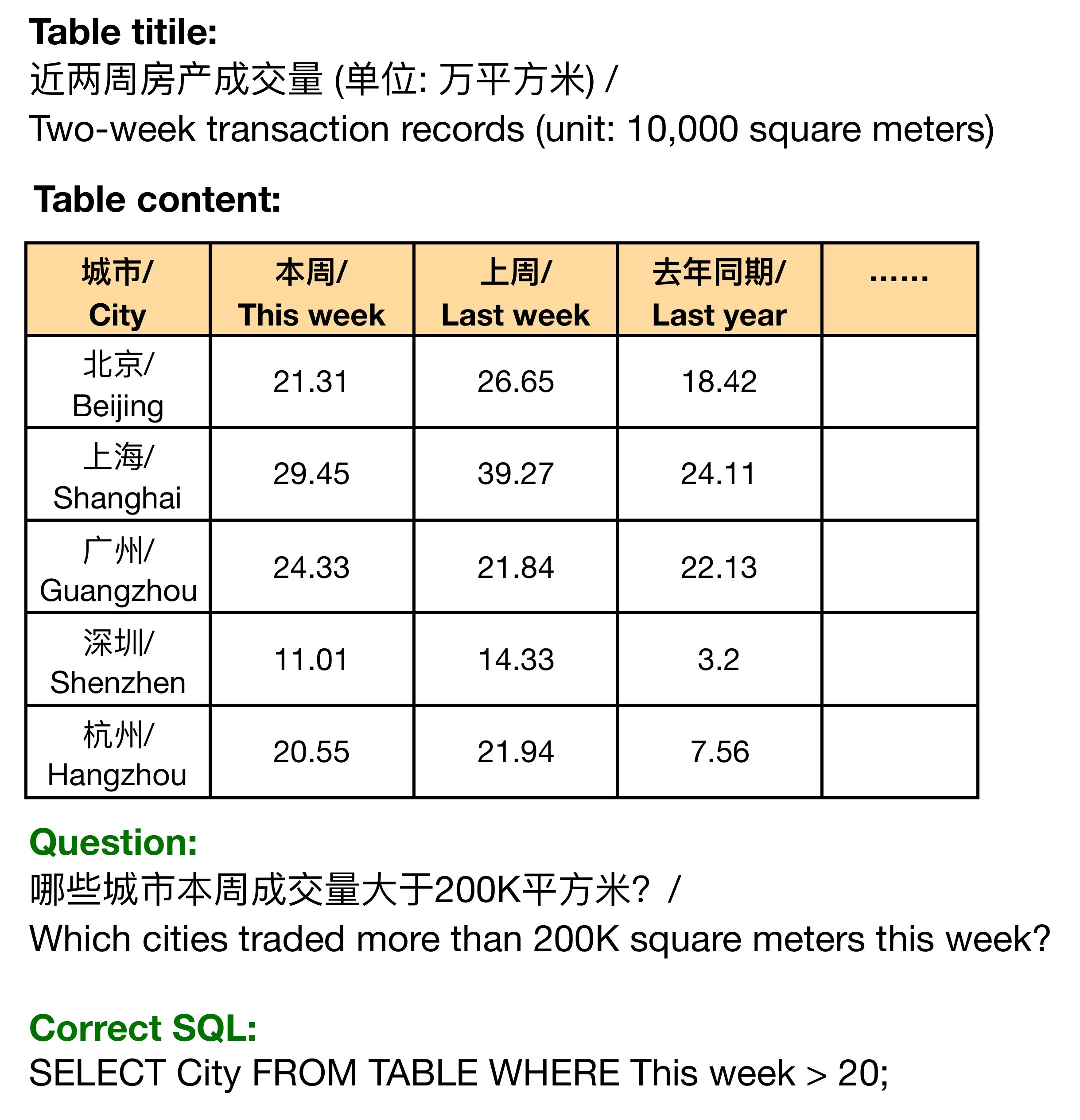}
    \caption{
    Sample data for numerical condition with inconsistent unit. 
    }
\label{numerical_case}
\end{figure}

\subsubsection{Complex SQL skeletons}
Table 2 summarizes the distribution of various SQL elements. Compared with WikiSQL, the TableQA presents a more challenging problem in the perspective of SQL skeletons. Since the models decouple the semantic parsing problem into a sequence of subtasks, the number of tasks increase when the supported SQL skeleton become more complex. So far, TableQA needs 9 subtasks, apparently, end-to-end models lose its advantages of keeping solution simple. 

\subsubsection{Error Accumulation} 
State-of-the-art models based on decoupled subtasks may suffer from the accumulation of errors. Take the structure of SQLova as shown in Figure \ref{model_structure} for example, in the condition prediction part, SQLova first predicts the number of conditions, then predicts condition columns and corresponding condition operations, and finally predicts condition values. It is a sequential model structure, and the performance may drop dramatically if any of these subtasks are not well solved. It is more reasonable to combine table content and question to infer the condition part jointly, instead of predicting these subtasks separately. 

\subsubsection{Numerical Term} 
There are various expressions for numerical conditions. Take the sample shown in Figure \ref{numerical_case} as example, users want to query city names conditioned on the trade volume of this week, they may ask \textit{``Which cities traded more than 200K square meters this week?"}, but the unit of the corresponding column is ten-thousand, leading to a unit inconsistency problem. In this case, to generate the correct SQL, a system needs to combine the units used in utterance and tables. 

\section{Conclusion}\label{conclusion}
This paper introduces a large-scale Chinese NL2SQL dataset named TableQA. Compared to existing NL2SQL datasets, TableQA has several unique features and propose challenges challenging for NL2SQL research, including entity linking and answerability. These problems are helpful to successful application of NL2SQL. Besides the challenging dataset, two table-aware solutions for the entity linking problem are tested, while experiment results empirically prove their effectiveness.

\bibliography{main}
\bibliographystyle{aaai}
\end{document}